\documentstyle[12pt]{article}
\begin{document}
\newcommand{\be}{\begin{equation}}
\newcommand{\ee}{\end{equation}}
\begin{center}

{\bf QCD and hybrid NBD on oscillating moments of multiplicity distributions
in lepton- and hadron-initiated reactions.}\\

\vspace{3mm}

I.M. Dremin\footnote{Email: dremin@lpi.ru}\\

Lebedev Physical Institute, Moscow 119991, Russia\\ 

\end{center}

\begin{abstract}
QCD predictions for moments of multiplicity distributions are compared 
with experimental data on $e^+e^-$ collisions and their two-NBD fits.
Moments of the multiplicity distribution in a two-NBD model for 1.8 TeV 
$pp$-collisions are considered. Three-NBD model predictions and fits 
for $pp$ at LHC energies are also discussed. Analytic expressions for 
moments of hybrid NBD are derived and used to get insight into jet 
parameters and multicomponent structure of the processes. Interpretation 
of observed correlations is proposed.
\end{abstract}

Multiplicity distributions are the integral characteristics of multiparticle 
production processes. They can be described either in terms of probabilities
$P_n(E)$ to create $n$ particles at energy $E$ or by the moments of 
these distributions. It has been found that their shapes possess some 
common features in all reactions studied. At comparatively low energies 
below tens of GeV, these distributions are relatively
narrow and have sub-Poissonian shapes. At energies about 20 GeV for $e^+e^-$-
annihilation and 30 GeV for $pp$ (and $p\bar p$) interactions, they can be well 
fitted by the Poisson distribution\footnote {This difference of thresholds can 
be attributed to lower energy spent in hadron collisions for new particles 
creation (due to the so-called leading particle effect) compared to $e^+e^-$-
annihilation.}. At higher energies, the shapes become super-Poissonian, i.e. 
their widths are larger than for Poisson distribution. They increase with 
energy and, moreover, some shoulder-like substructures appear. 

Their origin is usually ascribed to multicomponent contents of the process.
In QCD description of $e^+e^-$-processes these could be subjets formed inside 
quark and gluon jets (for the reviews see, e.g., \cite{koch, dgar}).
In phenomenological approaches, the multiplicity distribution in a single 
subjet is sometimes approximated by the negative binomial distribution (NBD)
first proposed for hadronic reactions in \cite{giov}. For hadron-initiated 
processes, these peculiarities are also 
explained by the multicomponent structure of the process. This is either
multiladder exchange in the dual parton model \cite{cstt, kaid}, varying number 
of clans \cite{gugo} or multiparton interactions \cite{alex, mwal}.
These subprocesses are related to the matter state during the collision
(e.g., there are speculations about nonhomogeneous matter distribution in
impact parameters \cite{bour}, not to speak of quark-gluon plasma \cite{mcle}
behaving as a liquid \cite{shur} etc).

Such evolution of the multiplicity distributions can be quantitatively 
described by the energy behaviour of their moments. These moments reveal
the correlations inherent for the matter state formed during the collision.
Similarly to virial coefficients in statistical physics they can tell us 
about the equation of state of this matter. To introduce them, let us
write the generating function of the multiplicity distribution as
\be
G(E,z) = \sum_{n=0}^{\infty }P_{n}(E)(1+z)^{n}.                     \label{3}
\ee
In what follows, we will use the so-called unnormalized factorial ${\cal F}_q$
and cumulant ${\cal K}_q$ moments defined according to the formulae
\be
{\cal F}_{q} = \sum_{n} P_{n}n(n-1)...(n-q+1) =
 \frac {d^{q}G(E,z)}{dz^{q}}\vline _{z=0}, 
\label{4}
\ee
\be
{\cal K}_{q} = \frac {d^{q}\ln G(E,z)}{dz^{q}}\vline _{z=0}. \label{5}
\ee
They correspondingly define the total and genuine correlations among the 
particles produced (for more details see \cite{ddki, dgar}). These
cumulant moments could be considered as the direct analogies of virial 
coefficients of statistical physics since both are related to
genuine (irreducible) correlations. In particular, 
the first moments describe the mean multiplicity $\langle n \rangle $
\be
{\cal F}_1={\cal K}_1=\langle n \rangle , \label{nav}
\ee
and the second moments are related to the dispersion $D$ of the distribution
$P_n$:
\be
{\cal K}_2={\cal F}_2-\langle n \rangle ^2=D^2-\langle n \rangle . \label{disp}
\ee
The higher rank moments reveal other asymmetries of distributions such as
skewness etc. Since both ${\cal F}_q$ and ${\cal K}_q$ strongly increase with
their rank and energy, their ratio
\be
H_q={\cal K}_q/{\cal F}_q  \label{hq}
\ee
first introduced in \cite{13} is especially useful due to partial 
cancellation of these dependences. The factorial moments ${\cal F}_q$'s are
always positive by definition (Eq. (\ref{4})) while the cumulant moments 
${\cal K}_q$'s can change sign.
Again, let us recall that the changing sign second virial coefficient
in statistical physics implies the liquid state with the Van der Waals
equation corresponding to repulsion at small distances and attraction
at large distances. Cooper pair formation is also related to similar
behaviour of correlations.

Here, we compare QCD and NBD approaches to the description of multiplicity
distributions. We argue that $H_q$ values are more sensitive to minute details 
of the distributions than their direct chi-square fits and reveal differences
between proposed fits of $e^+e^-$ and $pp \; (p\bar p)$ processes. Some 
estimates for LHC energies will be provided.

The generating functions for quark and gluon jets satisfy definite equations
in perturbative QCD (see \cite{dkmt, dgar}). It has been analytically 
predicted in gluodynamics \cite{13} that at asymptotically high energies 
the $H_q$ moments are positive and decrease as $q^{-2}$ but at present
energies they become negative at some values of $q$ and reveal the negative 
minimum at
\be
q_{min}=\frac {1}{h_1\gamma_0}+0.5+O(\gamma _0),  \label{qmin}
\ee
where $h_1=b/8N_c=11/24, \;\; b=11N_c/3-2n_f/3, \;\; 
\gamma _0^2=2N_c\alpha_S/\pi, \;\; \alpha _S$ is a coupling strength, 
$N_c, \; n_f$ are the numbers of colours and flavours. At $Z^0$ energy 
$\alpha _S\approx 0.12$, and this minimum is at about $q\approx 5$. It moves to 
higher ranks with energy increase because the coupling strength decreases. Some 
hints to possible oscillations of $H_q$ vs $q$ at higher ranks at LEP energies 
were obtained 
in \cite{13}. Then the approximate solution of the gluodynamics equation
for the generating function \cite{41} agreed with this and predicted the 
oscillating behaviour at higher ranks. These oscillations were confirmed
by experimental data for $e^+e^-$ and hadron-initiated processes first in
\cite{dabg}, later in \cite{sld} and most recently in \cite{l3}.
The same conclusions were obtained from exact solution of equations for
quark and gluon jets in the framework of fixed coupling QCD \cite{21}.
The physics interpretation of these oscillations as originating from
multisubjet structure of the process is related to the (multi)fractal
behaviour of factorial moments, found also in QCD 
\cite{68, 38, 70}. The asymptotic disappearance of oscillations can be 
ascribed to the extremely large number of subjets at very high energies.

A recent exact numerical solution of the gluodynamics equation in a wide 
energy interval \cite{bfoc} coincides with the qualitative features of
multiplicity distributions described above. In terms of moments they
correspond to the values of $H_q$ changing sign at each subsequent $q$
(with $H_2<0$) at low energies (narrow shapes\footnote{Narrow distributions 
always have such cumulants as shown, e.g., in \cite{dgar}.}), the approach 
of $H_q$ to zero at the Poisson transition point about 20 GeV for $e^+e^-$ 
processes, and the positive second moment $H_2$ with oscillations of
higher rank cumulants at $Z^0$ which disappear asymptotically. At $Z^0$, the
first minimum appears at $q\approx 5$. This confirms earlier exact QCD 
results \cite{lo1} at $Z^0$. It moves to higher ranks with a steadily 
decreasing amplitude when energy increases. 
The only free parameter is the QCD cut-off, which is however approximately 
fixed by the coupling strength and does not strongly influence the results.

In parallel, the NBD-fits of multiplicity distributions were attempted 
\cite{gugo, gug1}. The single NBD-parameterization is
\be
P_n(E)=\frac{\Gamma(n+k_1)}{\Gamma(n+1)\Gamma(k_1)}\left (\frac{n_1}{k_1}\right)^n
\left(1+\frac{n_1}{k_1}\right)^{-n-k_1},  \label{pnbd}
\ee
where $\Gamma $ denotes the gamma-function. This distribution has two adjustable 
parameters $n_1(E)$ and $k_1(E)$ which depend on energy. Such a formula happened 
to describe low energy data with negative values of $k_1$ that corresponds to 
binomial fits. At the Poisson transition point $k^{-1}=0$. The parameter $k_1$ 
becomes positive at higher energies. However the simple fit by the formula 
(\ref{pnbd}) is valid till the shoulders appear. In that case, this formula 
is replaced by the hybrid NBD which combines two or more expressions like 
(\ref{pnbd}). Each of them has its own energy dependent parameters $n_i, k_i$.
These distributions are weighted with the energy dependent probability factors 
$\alpha_i$ which sum up to 1. Correspondingly, the number of adjustable 
parameters drastically increases.

A single NBD (\ref{pnbd}) has positive cumulants for $k_1>0$ (${\cal K}_q =\Gamma (q)n_1^q/
k_1^{q-1}$) and thus the positive $H_q=\Gamma(q)\Gamma(k_1+1)/\Gamma(k_1+q)$.
For hybrid NBD, the negative $H_q$ can exist. The traditional procedure to
calculate higher rank moments is by the iterative relations
\be
H_q=1-\sum_{m=1}^{q-1}\frac{\Gamma(q)}{\Gamma(m+1)\Gamma(q-m)}H_{q-m}\frac
{{\cal F}_m{\cal F}_{q-m}}{{\cal F}_q}.   \label{hqfq}
\ee
The strong compensations are inherent in Eq. (\ref{hqfq}). This calls for high
accuracy of numerical calculations. More important, the formula does not give 
any direct insight into the physical reasons for such compensations. Therefore,
it is instructive to write the analytic formulae for moments of hybrid NBD
which provide clear interpretation of negative values of cumulants. We have
derived these expressions for the two-NBD parameterization (2NBD) given by a 
sum of two expressions like (\ref{pnbd}) with two sets of adjustable parameters
$n_1, k_1$,  $\; n_2, k_2$ weighted with energy dependent factors $\alpha $ 
and $1-\alpha $ correspondingly. 2NBD describes the process with two 
independent NBD-components of mean multiplicities $n_i$ and widths $k_i$ created 
with probabilities $\alpha $ and $1-\alpha $. The factorial moments for any 
rank $q$ are given by the simple formula 
\be
{\cal F}_q=\alpha \frac{\Gamma(k_1+q)}{\Gamma(k_1)}\frac {n_1^q}{k_1^q}+(1-
\alpha) \frac{\Gamma(k_2+q)}{\Gamma(k_2)}\frac {n_2^q}{k_2^q} \;\;\;\;\;\;\;\;
(0\leq \alpha \leq 1).     \label{fact}
\ee
The cumulant moments are more complicated and should be calculated separately 
for each rank. The first 5 moments are
\begin{eqnarray}
{\cal K}_1={\cal F}_1=\langle n\rangle =\alpha n_1+(1-\alpha )n_2, \hspace{93.6mm}\\ 
\label{kf1}
\vspace{1mm}
{\cal K}_2=\frac{\alpha n_1^2}{k_1}+\frac{(1-\alpha )n_2^2}{k_2}+\alpha (1-
\alpha )(n_1-n_2)^2, \hspace{74.6mm}     \\   
\label{k2}
\vspace{1mm}
{\cal K}_3=\frac{2\alpha n_1^3}{k_1^2}+\frac{2(1-\alpha )n_2^3}{k_2^2}+\alpha 
(1-\alpha )(n_1-n_2)[3(n_1^2/k_1-n_2^2/k_2)+ 
(1-2\alpha )(n_1-n_2)^2], \hspace{0.65mm}\\
\label{k3}
\vspace{1mm}
{\cal K}_4=\frac{6\alpha n_1^4}{k_1^3}+\frac{6(1-\alpha )n_2^4}{k_2^3}+\alpha 
(1-\alpha )[(n_1-n_2)^4(1-6\alpha(1-\alpha ))+ 11(n_1^2/k_1-n_2^2/k_2)-
\nonumber \\     
8n_1n_2
(n_1/k_1-n_2/k_2)^2+6(1-2\alpha )(n_1-n_2)^2(n_1^2/k_1-n_2^2/k_2)],\hspace{0.6mm} \\ 
\label{k4}
\vspace{1mm}
{\cal K}_5=\frac{24\alpha n_1^5}{k_1^4}+\frac{24(1-\alpha )n_2^5}{k_2^4}+5\alpha 
(1-\alpha )[6(n_1-n_2)(n_1^4/k_1^3-n_2^4/k_2^3)+\hspace{30.3mm} \nonumber  \\
4(n_1^3/k_1^2-n_2^3/k_2^2)
(n_1^2/k_1-n_2^2/k_2)+(1-2\alpha )(n_1-n_2)(7(n_1-n_2)(n_1^3/k_1^2-n_2^3/k_2^2)
+\nonumber \\
3n_1n_2(n_1/k_1-n_2/k_2)^2)+2(1-6\alpha(1-\alpha))(n_1-n_2)^3
(n_1^2/k_1-n_2^2/k_2)+ \nonumber  \\
0.2(1-2\alpha )(1-12\alpha (1-\alpha ))(n_1-n_2)^5]
, 
\label{k5} 
\end{eqnarray}

For $\alpha=0$ or 1 they reduce to one-NBD formulae with one of the first two 
terms surviving. It is always positive for positive $k_i$. Therefore, as 
expected, individually considered, the distributions show no oscillations. For 
2NBD, there is a symmetry in replacing indices 1 to 2 together with $\alpha $
to $1-\alpha $. Negative ${\cal K}_2$ can be obtained only if $k_i<0$. For
positive $k_i$ one always gets positive ${\cal K}_2$. Its value depends on the 
difference $n_1-n_2$. ${\cal K}_3$ can become negative depending on the values 
of the last two terms. The cancellations in the expressions (\ref{k2})-(\ref{k5})
are not so drastic as in Eq. (\ref{hqfq}), especially for large $q$, because 
the leading contributions to $H_q$ are strongly decreasing with $q$ there and
not of the order of 1 as in (\ref{hqfq}). Therefore 
they do not require very high precision and, moreover, clearly display the 
origin of each term and its dependence on fitted parameters.

Actually, five moments determine quite well the shape of the distribution if 
they are calculated with high enough accuracy. Since these shapes are 
qualitatively similar in different reactions, it is especially instructive to 
compare their $H_q$ moments. In Table 1 the $H_q$ moments for $e^+e^-$ 
annihilation at $Z^0$ are shown. Their values according to the solution of 
the gluodynamics equations \cite{bfoc} are in the first column. In the second
and third columns, the experimental results of L3 collaboration \cite{l3}
are represented for full phase space (FPS), correspondingly, with all 
measured multiplicities included and with some very high multiplicities 
truncated (because of large error bars). Next follow $H_q$ values restored
from 2NBD fits of OPAL and DELPHI results done in \cite{glug}.\\ 

{\bf Table 1.}  \\

\begin{tabular}{|c|c|c|c|c|c|} \hline
 &QCD & L3, untr& L3, tr & OPAL,2NBD & DELPHI,2NBD  \\ \hline
$H_2$& 3.9E-2& (4.42$\pm$0.11)E-2& (4.41$\pm$0.10)E-2& 4.4E-2& 3.1E-2  \\
$H_3$ & 7.4E-3& (7.40$\pm$0.38)E-3 & (7.20$\pm$0.35)E-3& 7.4E-3& 2.6E-3 \\
$H_4$ & 4.0E-4& (9.69$\pm$2.56)E-4& (7.17$\pm$1.42)E-4& 4.9E-4 & 7.4E-5 \\
$H_5$ & -2.2E-4& -(1.30$\pm$1.59)E-4& -(3.95$\pm$0.53)E-4& -2.4E-4&-7.3E-5 \\
\hline
\end{tabular}\\

The overall agreement is rather good. Quite impressive is the fact that in
all cases the fifth cumulant moment is negative. However, somewhat
surprising is the difference of the theoretical and experimental widths
($H_2$ values). The widths are determined quite precisely both experimentally
and theoretically. The only reason to which such disagreement could be ascribed
is the incomplete treatment with quarks omitted in \cite{bfoc}. The more
complete approach will shed more light on this problem. The even stronger
disagreement with DELPHI data is probably related to the special selection of 
events there. Further analysis is needed.

The comparison of $e^+e^-$ and $pp\;(p\bar p)$ data turns out especially
interesting. While both show qualitative similarity of the shapes of 
multiplicity distributions, the corresponding $H_q$ values are quite 
distinctive. In Table 2 we show $H_q$ values for $p\bar p$ at 1.8 TeV 
(Tevatron) and interpolations to 14 TeV (LHC). The 2NBD fit at 1.8 TeV
corresponds to the following parameters: $\alpha=0.62, \; n_1=30, \; n_2=61.6,
\; k_1=k_2=7$, which are approximately equal to average values for 2.A model 
considered in \cite{gugo}. However, even this extreme model underestimates
high multiplicities and, therefore, $H_q$ values in the Table should be treated
as lower bounds to experimental ones, which are unknown, unfortunately.
The extrapolated values at 14 TeV have been calculated
using the parameters of 3NBD fits and Pythia model considered in \cite{gug1}.\\

{\bf Table 2.}\\

\begin{tabular}{|c|c|c|c|c|} \hline
 &2NBD fit, 1.8TeV& 3NBD fit, 14TeV & Pythia, 14TeV  \\ \hline
$H_2$ &0.2279 &  0.8754& 0.4224  \\
$H_3$  & 0.0988  & 0.9703& 0.3387 \\
$H_4$  &0.0414  &0.9737  & 0.2683 \\
$H_5$  &0.0120  &0.9742  &0.1877 \\
\hline
\end{tabular}\\

Quite impressive are much larger values of $H_q$ in hadron-initiated reactions
(Table 2) as compared to $e^+e^-$ results (Table 1). They strongly increase with 
energy. Moreover, the drastic difference is clearly displayed by $H_q$ 
between 3NBD interpolations and Pythia at 14 TeV.
This demonstrates extremely high sensitivity of $H_q$ analysis because both 
approaches provide the similar two-shoulder structure
of multiplicity distributions as seen in Fig. 2 of \cite{gugo}. 
At 14 TeV, the predictions are given for full phase space. For the 
rapidity interval $\vert \eta\vert<0.9$ the $H_q$ values become larger than
those in Table 2. $H_q$ for the 3NBD-model of \cite{gugo} become almost 
indistinguishable from 1 (above 0.99). Pythia values increase about 1.4 times.
No oscillations are seen at these high energies while they are present at 
energies below 1 TeV \cite{dabg}. 
Surely, LHC experiments will give their decisive conclusion.

To conclude, we have shown that $H_q$ moments of the multiplicity distribution 
are extremely sensitive to minute details of its shape. They can resolve the 
differences between various fits even if those are not clearly seen in the 
traditional representation. For $e^+e^-$, slight disagreement on theoretical and 
experimental widths is embarassing and must be further studied. For hadron- and
nuclei-initiated reactions, $H_q$ values are much larger than in $e^+e^-$. 
RHIC and LHC data are awaited for better insight. The energy dependence of 
$H_q$ and of the relative weights of various NBD components can provide some 
hints on the matter state during the collision and its energy evolution.

I am grateful to A. Giovannini and W. Metzger for correspondence.
This work has been supported in part by the RFBR grants N 02-02-16779,
03-02-16134, NSH-1936.2003.2.


\begin{thebibliography}{99}
\bibitem{koch}
V.A. Khoze and W. Ochs, Int. J. Mod. Phys. A  12 (1997) 2949.
\bibitem{dgar}
I.M. Dremin and J.W. Gary, Phys. Rep. 349 (2001) 301.
\bibitem{giov}
A. Giovannini, Nuovo Cim. A 15 (1973) 543. 
\bibitem{cstt}
A. Capella, U. Sukhatme, C.I. Tan and J. Tran Thanh Van, Phys. Lett. B 81 
(1979) 68.
\bibitem{kaid}
A.B. Kaidalov, Phys. Lett. B 116 (1982) 459.
\bibitem{gugo}
A. Giovannini and R. Ugoccioni, Phys. Rev. D 59 (1999) 094020.
\bibitem{alex}
T. Alexopoulos et al, Phys. Lett. B 435 (1998) 453.
\bibitem{mwal}
S. Matinyan and W.D. Walker, Phys. Rev. D 59 (1999) 034022.
\bibitem{bour}
C. Bourelly et al, Proc. of the VI Blois Workshop on Frontiers in Strong 
Interactions, Ed. J. Tran Thanh Van, Editions Frontiers (1995), p.15.
\bibitem{mcle}
L. McLerran, Rev. Mod. Phys. 58 (1986) 1021.
\bibitem{shur}
E. Shuryak, hep-ph/0312227.
\bibitem{ddki}
E. DeWolf, I.M. Dremin and W. Kittel, Phys. Rep. 270 (1996) 1.
\bibitem{13}
I.M. Dremin, Phys. Lett. B 313 (1993) 209.   
\bibitem{dkmt}
Yu.L. Dokshitzer, V.A. Khoze, A.H. Mueller and S.I. Troyan, {\it Basics of
perturbative QCD} ed. by J. Tran Thanh Van
(Gif-sur-Yvette, Editions Frontieres, 1991).
\bibitem{41}
I.M. Dremin and V.A. Nechitailo, Mod. Phys. Lett. A  9 (1994) 1471;
JETP Lett. 58 (1993) 881.
\bibitem{dabg}
I.M. Dremin, V. Arena, G. Boca et al, Phys. Lett. B 336 (1994) 119.
\bibitem{sld}
SLD Collaboration, K. Abe et al, Phys. Lett. B 371 (1996) 149.
\bibitem{l3}
L3 Collaboration, P. Achard et al, Phys. Lett. B 577 (2003) 109.
\bibitem{21}
I.M. Dremin and R.C. Hwa, Phys. Rev. D 49 (1994) 5805; Phys. Lett. 
B 324 (1994) 477.
\bibitem{68}
W. Ochs and J. Wosiek,  Phys. Lett. B 289 (1992) 159; 
304 (1993) 144.
\bibitem{38}
Yu. L. Dokshitzer and I.M. Dremin, Nucl. Phys. B 402 (1993) 139.
\bibitem{70}
Ph. Brax, J.L. Meunier and R. Peschanski, Z. Phys. C 62 (1994) 649.
\bibitem{bfoc}
M.A. Buican, C. F\"{o}rster and W. Ochs, hep-ph/0307234.
\bibitem{lo1}
S. Lupia and W. Ochs, Phys. Lett. B 418 (1998) 214; Nucl. Phys.
(Proc. Suppl.) B 64 (1998) 74.
\bibitem{gug1}
A. Giovannini and R. Ugoccioni, hep-ph/0312205.
\bibitem{glug}
A. Giovannini, S. Lupia and R. Ugoccioni, Phys. Lett. B 374 (1996) 231.

\end{thebibliography}
\end{document}